\documentclass{elsart}

\usepackage{graphics}
\usepackage{graphicx}
\usepackage{epsfig}

\usepackage{amssymb}

\usepackage[english,francais]{babel}

\newtheorem{e-proposition}[theorem]{Proposition}

\newtheorem{e-definition}[theorem]{Definition\rm}

\newcommand{\un}{u^{(n)}}
\newcommand{\vn}{v^{(n)}}
\newcommand{\rhon}{\rho^{(n)}} 
\newcommand{\psin}{\psi^{(n)}} 
 
\newcommand{\unmean}{\langle u^{(n)} \rangle} 
\newcommand{\vnmean}{\langle v^{(n)} \rangle}

\def\og{\leavevmode\raise.3ex\hbox{$\scriptscriptstyle\langle\!\langle$~}}
\def\fg{\leavevmode\raise.3ex\hbox{~$\!\scriptscriptstyle\,\rangle\!\rangle$}}

\begin{document}

\begin{frontmatter}


\selectlanguage{english}
\title{Wave turbulence and Bose-Einstein condensates}

\vspace{-2.6cm}

\selectlanguage{francais}
\title{Fluctuations turbulentes dans les condensats de Bose-Einstein}


\selectlanguage{english}
\author[Joss]{Christophe Josserand}
\ead{josseran@lmm.jussieu.fr}

\address[Joss]{Laboratoire de Mod\'elisation en M\'ecanique, 
Universit\'e Pierre et Marie Curie and CNRS\\
Case 162, 4 place Jussieu, 75252 Paris Cedex 05, France}

\begin{abstract}{
Asymptotic behavior of a class of nonlinear Schr\"odinger equations are 
studied. Particular cases of 1D weakly focusing and Bose-Einstein 
condensates are considered. A statistical approach is presented following\cite{JJ00} to 
describe the stationary probability density of a discretized finite
system. Using a maximum entropy argument, the theory predicts that the
statistical equilibrium is described by energy equivalued fluctuation modes
around the coherent structure minimizing the Hamiltonian of the system. 
Good quantitative agreement is found with numerical simulations.
In particular, the particle number spectral density follows an effective 
$1/k^2$ law for the asymptotic large time averaged solutions. Transient
dynamics from a given initial condition to the statistically steady regime
shows rapid oscillation of the condensate.
\vskip 0.5\baselineskip

\selectlanguage{francais}
\noindent{\bf R\'esum\'e}
\vskip 0.5\baselineskip
\noindent
Le comportement asymptotique des solutions d'\'equations diff\'erentielles 
hamiltoniennes est pr\'esent\'e dans le cas g\'en\'eral des \'equations de 
Schr\"odinger nonlin\'eaires. Ce travail reprend une \'etude pr\'ec\'edente
s'appuyant sur une description statistique de l'espace des phases de la
solution\cite{JJ00}. La recherche de la distribution stationnaire 
\`a l'\'equilibre statistique s'effectue pour la dynamique discr\`ete en 
maximisant l'entropie autour de la solution concentrant toute la masse du
syst\`eme. On trouve alors que la distribution d'\'equilibre correspond
\`a l'\'equipartition statistique de l'\'energie en exc\`es sur tous les
modes accessibles. Les simulations num\'eriques sur un mod\`eles faiblement
focalisant et dans le cas particulier d'un mod\`ele 1D de condensat de 
Bose-Einstein permettent de montrer un bon accord quantitatif avec les
pr\'edictions de la th\'eorie.
}

\vskip 0.5\baselineskip
\noindent{\small{\it Mots-cl\'es~: turbulence d'ondes, condensats, distribution d'\'equilibre}}}
\end{abstract}
\end{frontmatter}

\selectlanguage{francais}
\section*{Version fran\c{c}aise abr\'eg\'ee}

\selectlanguage{english}

\section{Introduction}
\label{}

The emergence and persistence of large scale coherent structures in the
midst of small scale turbulent fluctuations is a common feature
of many turbulent fluids, plasma systems and nonlinear dynamics in 
general\cite{tab,Mcwilliams,HK,RN01}. Classical examples include the formation of 
large vortices in two dimensional high Reynolds turbulence or the emergence of
solitary solutions in nonlinear optics\cite{vort,Pesceli}. Moreover and 
quite surprisingly, such complex phenomena survive even when the 
dissipationless versions of the dynamics are taken. In these cases as in 
many classical Hamiltonian systems, the dynamics are formally 
reversible\cite{hasa}. Therefore one can question how such apparently 
irreversible processes can be compatible with these irreversible dynamics?
The recent attainment of Bose-Einstein 
condensates (BEC)\cite{Dav95,And95,BEC} provides a fascinating new system 
exhibiting similar behavior. After being attained, the condensate can indeed 
be viewed as a large coherent structure, persisting in the atomic trap and 
surrounded by fluctuations. When thermal fluctuations are neglected, BEC 
can be modelled by the usual semi-classical Gross-Pitaevsk\u{\i}i
equation. It is strictly valid only at $T=0$ and is also Hamiltonian 
dynamics. Moreover note that for all these examples the dynamics resume in 
partial differential equations describing the evolution of one or few classical
fields (velocity for turbulent flows or wave function for BEC for instance).\\

The interplay between the fluctuations and the coherent structures in these
systems is of crucial interest to our understanding of nonlinear dynamics. 
Important questions are related to the asymptotic behavior of the dynamics
and to their possible statistical description. In particular, the expected
thermal equilibrium would in fact lead to the well-known Rayleigh-Jeans 
divergences for a classical field. The aim of this note is precisely to 
investigate how this black body-like catastrophy manifests in the {\it a 
priori} smooth dynamics driven by the partial differential equations (PDE)
considered here. 
For simplified reversible models, it is believed and borne out by numerical
simulations that the 
coherent structures may act as statistical attractors to which the whole 
system relaxes. Following weak turbulence theory, in 
continuum systems, the fluctuations
were shown to decrease without bound in addition to cascading to smaller and
smaller scales. The explanations proposed for this cascade-like process invoke 
in fact thermodynamical considerations\cite{ZPSY,Pomeau}. In collaboration
with R. Jordan, we have recently focused on such scenario for the
class of nonintegrable nonlinear Schr\"odinger (NLS) 
equations\cite{JJ00}. Following \cite{JTZ} we seek the probability 
density of a solution for a finite number $n$ of modes version of the NLS 
equations. Usual thermodynamical arguments, where the probability density
corresponds to the standard Maxwell-Boltzmann distribution, fail and one has
to consider that a coherent structure emerges from the solution of the
PDE. 
Indeed, from numerical simulations we assume the presence of a 
non-zero mean field which contains most of the conserved particle number 
($L^2$-norm squared). Such specific treatment of a particular part of the
solution is therefore very close to the statistical theory of the Bose-Einstein
transition! The stationary probability distribution is obtained 
{\it via} the maximization of the entropy of this finite statistical system. 
It describes an ensemble where the mean-field corresponds to a large--scale 
coherent solitary wave, which minimizes the Hamiltonian given the particle 
number, coupled with small--scale random fluctuations, or radiation.  The
fluctuations equally share the difference of the conserved 
value of the Hamiltonian and the Hamiltonian of the coherent state. The 
effective temperature of this thermal-like system is inversely proportionnal 
to $n$ the number of modes and goes therefore to zero in the continuum limit. 
Thus the discretization level $n$ of the dynamics triggers an effective 
energy cut-off which avoids the Rayleigh-Jeans divergencies. This
statistical theory found very good qualitative and quantitative agreement
with numerical simulations done for a weakly focusing NLS equation.
Statistical ensemble can be retrieved by time and ensemble averaging of 
large time numerical solutions starting from random initial conditions.
At large enough time, the dynamics of the solutions present statistical
stationary properties in full agreement with the statistical theory.
The road to this statistically stationary state is also investigated and 
allows a consistent scenario for the dynamics in the continuum limit 
$n \rightarrow \infty$.\\
In this note, I discuss how these results apply to the case of BEC 
dynamics. 
The Gross-Pitaevsk\u{\i} equation in harmonic trap used to model the condensate
evolution is in fact a particular case of NLS system so that the introduced 
description is valid. 
Before that, I will present in detail the general 
theory for the NLS equation in one spatial dimension with a finite number 
of modes.

\section{Self-organization in NLS-systems}

In this section I briefly introduce the results obtained for the 
NLS-equations. Most of the conclusions and the figures 
are presented for a specific $1$-D focusing NLS equation but the results
apply to the whole class of these models. The presentation here will follow
\cite{JJ00} and the reader will find more details and references there.

\subsection{Generalities}

We consider the general dimensionless NLS equation:

\begin{equation} \label{nls1} 
i \partial_t \psi + \Delta \psi + f(| \psi |^2)\psi\ = 0\,, 
\end{equation} 
where $\psi({\bf r},t)$ is a complex field and $\Delta$ is the 
Laplacian operator. The function $ f(|\psi|^2)$ stands for nonlinear
interaction and external potential.
Among other phenomena, it is used to model gravity waves on deep 
water\cite{AS}, Langmuir 
waves in plasmas \cite{Pesceli}, pulse propagation along optical fibers 
\cite{HK}, superfluid dynamics\cite{GP} and Bose-Einstein 
condensates\cite{DGPS99}. In this latter case, the interaction function
$ f(| \psi |^2)$ depends also on the position ${\bf r}$ to describe the 
atomic trap in which Bose-Einstein condensates are achieved. 
When $f(|\psi|^2)= \pm |\psi|^2$ and eqn.~(\ref{nls1}) is posed on the 
whole real line or on a bounded interval with periodic boundary 
conditions, the equation is completely integrable 
\cite{ZS}. In any other configuration it is nonintegrable. 
 
The Hamiltonian equation associated to (\ref{nls1}) is: $i 
\partial_t \psi 
= \delta H/ \delta \psi^*$, where $\psi^*$ is the complex conjugate of the 
field $\psi$, and $H$ is the Hamiltonian: 

\begin{equation} \label{ham1} 
H(\psi) = \int\left ( |{\bf \nabla} \psi|^2-F(|\psi|^2) \right 
)\,d{\bf r}\,. 
\end{equation}

Here, in addition to the kinetic term $|{\bf \nabla} \psi|^2$, the 
{\em potential} $F$ is defined via the relation $F(a) = \int_0^a 
f(y)\, dy$. The dynamics (\ref{nls1}) conserves, in addition to the 
Hamiltonian, the particle number (also called the mass)

\begin{equation} \label{part1} 
N(\psi)= \int |\psi|^2\, d{\bf r}\,. 
\end{equation}

Without loss of generality we hereafter restrict the statistical analysis 
and the numerics to non-integrable models in one spatial dimension.
Equations (\ref{nls1}) exhibit solitary wave solutions $ \psi=\phi(x,t)
e^{\imath \lambda^2 t} $ which satisfy

\begin{equation} \label{gse} 
\phi_{xx} + f(|\phi|^2)\phi - \lambda^2 \phi = 0\,. 
\end{equation} 

\begin{figure}
\centerline{  \epsfxsize=10truecm \epsfbox{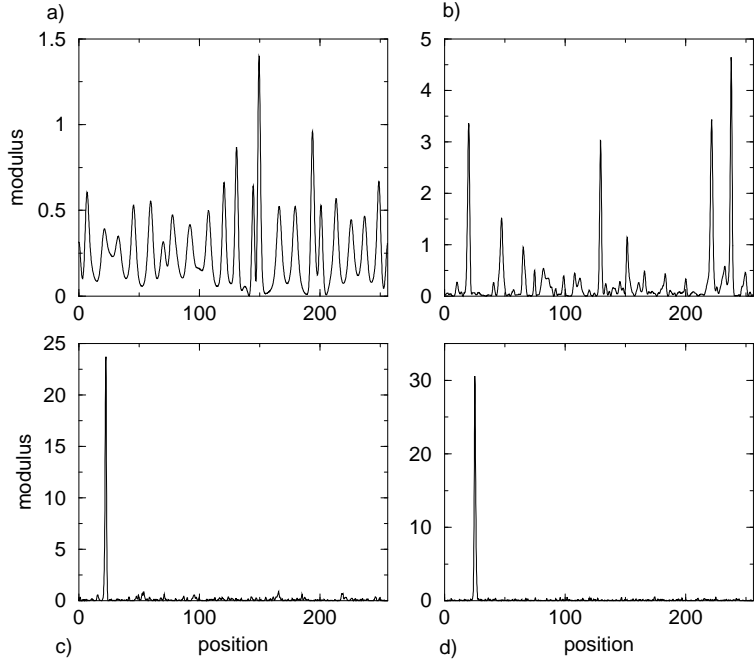}}  
\caption{\protect\small Profile of the modulus 
$|\psi|^2$ at four different times for the system (\ref{nls1})
with nonlinearity $f(|\psi|^2) = |\psi|$ and periodic boundary 
conditions on the
interval
$[0,256]$.   The initial 
condition is  $\psi(x, t=0 ) = A$, with $A = 0.5$,  plus
a small random perturbation.  The numerical scheme used to approximate
the solution is the split-step Fourier method. 
The  grid size  is $dx=0.125$, and the number of  modes is $n=2048$. 
a) $t=50$ unit time: Due to the modulational instability, an array of 
soliton-like structures separated by the typical distance  
$l_i=2\pi/\sqrt{A/2}=4\pi$ is created; b) $t=1050$ unit time: 
The solitons interact and coalesce, giving rise to a smaller number  
of solitons of larger amplitude; c)  $t=15050$:  The  
coarsening process has ended.  One large soliton  
remains in a background of  small--amplitude radiation. Notice that for  
$t=55050$ unit time (Figure d)), the amplitude of the fluctuations  
has diminished while the amplitude of the soliton has increased. 
\label{evolu}} 
\end{figure}

It is worthwhile to notice that the 
solution of equation (\ref{gse}) minimizes the Hamiltonian
(\ref{ham1}) for a given particle number. We will denote $ H^*(N)$ the
value of the energy of this solution. These localized structures are found
to play an important role in the time evolution of Eq. (\ref{nls1}).
Figure (\ref{evolu}) shows indeed snapshot of the dynamics for the weakly 
focusing 
nonlinearity $ f(|\psi|^2)=|\psi|$. Starting with a slightly perturbed
homogenous solution, a collection of solitary peaks emerge from this linearly 
unstable state. These solitary waves rapidly coalesce into a 
single coherent structure surrounded by small amplitude
fluctuations. A slow, quasistatic dynamics follows where the fluctuations are
seen to decrease in wavelength and in amplitude. The solitary structure gathers
almost all the mass of the system while the energy reaches smaller scales.
In other words, the dynamics presents the {\it condensation} of the mass into
a coherent structure while the energy is distributed in the system.

\subsection{Statistical model}

To explain these numerical observations, we construct a consistent 
mean--field statistical theory based on \cite{JTZ}.
It seeks to describe the solution as a mean field which contains
most of the particle number of the system while the amount of energy not
accounted by this solution is dispersed towards small scales.
The theory is in fact presented for finite dimensional approximation of the
NLS equation, using also Dirichlet boundary conditions on the interval 
$[0,L]$. Other boundary conditions, such as the periodic ones used in the 
simulations can easily be considered without any changes in the conclusions.

Let $e_j(x)=\sqrt{2/L}\sin(k_j x)$ with $k_j=\pi j/L$, and for any 
function $g(x)$ on $\Omega=[0,L]$ denote by $g_j = \int_{\Omega} g(x)e_j(x)\,
dx$ 
its $j$th Fourier coefficient with respect to the orthonormal basis $e_j, 
 j= 1,2,\cdots$. We consider now a truncated version of the NLS equations
to the $n$ first Dirichlet modes. We write $\psin = \un + i\vn$, where $\un$ and
$\vn$ are two real functions whose Fourier coefficients satisfy the coupled 
system of  
ordinary differential equations 
 
\begin{eqnarray} 
\dot{u}_j - k_j^2 v_j +  
\left ( f( (u^{(n)})^2 + (v^{(n)})^2) v^{(n)} \right )_j = 0
\\ 
\dot{v}_j + k_j^2 u_j -  
\left ( f( (u^{(n)})^2 + (v^{(n)})^2) u^{(n)} \right )_j = 0\,. 
\label{spectrunc} 
\end{eqnarray} 

It corresponds for $\psin$ to:

\begin{displaymath}
 i \psin_t + \psin_{xx} + P^n ( f(|\psin|^2) \psin ) = 0\,,
\end{displaymath}

where $P^n$ is the projection onto the span of the eigenfunctions 
$e_1, \cdots, e_n$. This equation is a natural spectral approximation 
of the NLS equation (\ref{nls1}), and it may be shown that its 
solutions  converge as $n \rightarrow \infty$ to solutions of 
(\ref{nls1}) \cite{Bourgain,Zhidkov}. 
 
For given $n$, the system of  equations (\ref{spectrunc})  
defines a dynamics on the $2n$--dimensional phase space ${\bf R}^{2n}$.
This  
finite-dimensional dynamical system is a Hamiltonian system, with  
conjugate variables $u_j$ and $v_j$, and with Hamiltonian 
 
\begin{equation} \label{disham} 
H_n = K_n + \Theta_n\,, 
\end{equation} 
where 
\begin{equation}\label{kn} 
K_n = \frac{1}{2} \int ((u^{(n)}_x)^2 + (v^{(n)}_x)^2)\, dx = 
 \frac{1}{2} \sum_{j=1}^n k_j^2 (u_j^2 + v_j^2)\,, 
\end{equation} 
is the kinetic energy, and 
\begin{equation} \label{tn} 
\Theta_n = -\frac{1}{2}\int F((u^{(n)})^2 + (v^{(n)})^2)\\ 
dx\,, 
\end{equation}  
is the potential energy. 
The Hamiltonian $H_n$ is, of course, an invariant of the dynamics.  
The truncated version of the particle number, up to a multiplicative factor in
the definition 
 
\begin{equation}\label{dispn} 
N_n = \frac12 \int ( (u^{(n)})^2 + (v^{(n)})^2)\, dx = \frac12 \sum_{j=1}^n (u_j^2 + v_j^2)\,, 
\end{equation} 
is also conserved by the dynamics (\ref{spectrunc}).

We argue that we can build a statistical treatment of the system using the 
usual assumption that the dynamics is ergodic and noting also that 
system (\ref{spectrunc}) satisfies the Liouville property (the measure 
 $\prod_{j=1}^n du_j dv_j$ is invariant)\cite{balescu}.\\
We thus replace the time dependant dynamics by a statistical (time independant)
description of the solution at any time. We introduce a probability density 
$\rhon(u_1, \cdots, u_n, v_1\, \cdots, v_n)$ 
on the $2n$--dimensional phase space and we seek the density 
function $\rhon$ which maximizes the Gibbs-Boltzmann entropy functional:

\begin{equation}\label{gbent} 
S(\rho) = -\int_{{\bf{R}}^{2n}} \rho \log \rho  
\prod_{j=1}^n du_j dv_j\,, 
\end{equation}

We easily obtain the usual canonical ensemble solution:

\begin{displaymath}
\rho \propto
\exp \left ( -\beta H_n - \mu N_n \right )\,,
\end{displaymath}

subject to the mean constraints $\langle H_n \rangle=H^0$ and  
$\langle N_n \rangle=N^0$. $H^0$ and $N^0$ are the given values of
the Hamiltonian and the particle number, respectively, and
$\beta$ and $\mu$ are the Lagrange multipliers associated to these
constraints. Such a density is actually ill--defined since it is not 
normalizable  (i.e.,
$\int_{{\bf R}^{2n}} \exp [ -\beta H_n - \mu N_n ]
\prod_{j=1}^n du_j dv_j$ diverges) for the focusing nonlinearities we discuss 
here\cite{JTZ,LRS}. Moreover, this distribution does not take into account the numerical observed fact that an 
important part of the phase space consists of configurations where
most of the mass is concentrated in the solitary wave solution.\\
Inspired by these remarks, we have built an adapted statistical description of
the dynamics. We decompose the fields $ u_n$ and $ v_n$ into two contributions:
the means denoted $\langle u_j \rangle$ and $\langle v_j \rangle$ and the
fluctuations $(\delta \un, \delta \vn) \equiv (u^{(n)}-\unmean, v^{(n)}-
\vnmean)$. This decomposition will be clarified by the yet to be determined 
ensemble $\rhon$. We now impose that in the long-time NLS dynamics 
(\ref{spectrunc}) and for the continuum limit $ n \rightarrow \infty$, the
amplitude of the fluctuations vanishes. Consequently, the number of particles
in this limit is almost determined by the mean field. The vanishing
fluctuations hypothesis is written:

\begin{equation} \label{vfh}  
\int_{\Omega} \left [ \langle (\delta \un )^2 \rangle +  
\langle (\delta \vn )^2\rangle \right ] dx \equiv 
\sum_{j=1}^n \left [ \langle (\delta u_j)^2 \rangle +   
\langle (\delta v_j)^2
\rangle \right ] 
\rightarrow 0\,,\; \mbox{as }\; n \rightarrow\infty\,. 
\end{equation} 

Using this assumption, it follows that for $n$ large enough, the total
number of particles is well approximated by the mean contribution.
Similarly, one can show that the potential energy is almost entirely
determined by the potential of the mean. However, although the fluctuations
do not contribute much to the particle number and the potential energy, they
may have a significant kinetic energy.  Indeed, this contribution $(1/2) 
\sum_{j=1}^n k_j^2 [ \langle (\delta u_j)^2 \rangle  
+\langle (\delta v_j)^2 \rangle ]$, need not tend to 0 as $n 
\rightarrow \infty$, even if (\ref{vfh}) holds. Actually, it must in 
general not tend to $0$ since the total energy of the system is conserved.

From these remarks, we can impose the following mean-field constraints on 
$ \rho$:

\begin{eqnarray}\label{mfc} 
\tilde{N}_n(\rho) \equiv \frac{1}{2} \sum_{j=1}^n (\langle u_j \rangle^2 
+ \langle v_j \rangle^2) = N^0  \\ 
\tilde{H}_n (\rho) \equiv \frac{1}{2} \sum_{j=1}^n k_j^2 (\langle 
u_j^2 \rangle + \langle v_j^2 \rangle)  
-\frac{1}{2} \int_{\Omega} F(\langle u^{(n)} \rangle^2 + \langle v^{(n)} 
\rangle^2)\,dx = H^0\,. 
\end{eqnarray} 

Where $ N^0$ and $ H^0$ are precisely the conserved values of the particle 
number and the energy, as determined from initial conditions.

The solutions $ \rhon$ of the statistical equilibrium states with the
particle number constraint (\ref{mfc}) are calculated now invoking again a maximum 
entropy principle. Notice that, similar to its above use, this principle has 
no reason to hold in such hamiltonian and reversible systems. However, it 
allows a consistent calculation of the density following an ergodic 
assumption\cite{balescu}. It somehow corresponds to the determination of
the density distribution around the most probable state.\\

The solutions $ \rhon$ are therefore calculated through usual techniques: two
Lagrange multipliers are used to enforce the mass and the energy constraints.
Considering independent statistical variables, the factorization of the 
maximum entropy distribution is straightforward:

\begin{equation}\label{rho1} 
\rho^{(n)}(u_1, \ldots, u_n, v_1, \ldots, v_n) = \prod_{j=1}^n \rho_j(u_j, 
v_j)\,, 
\end{equation} 
where, for $j=1,\ldots, n$,  
 
\begin{equation}\label{rho2} 
\rho_j(u_j,v_j) = \frac{\beta k_j^2}{2 \pi} \exp \left \{ -\frac{\beta 
k_j^2}{2}\left ((u_j - \langle u_j \rangle)^2 
+ (v_j - \langle v_j \rangle)^2 \right ) \right \}\,, 
\end{equation} 

In addition, we find that the complex field  $\langle \psi^{(n)} \rangle =  
\unmean + i \vnmean$ is a solution of (setting $\lambda= \mu/\beta$) 
\begin{equation}\label{rho} 
\langle \psi^{(n)} \rangle_{xx}+P^n \left (f(|\langle \psi^{(n)} 
  \rangle|^2)\langle \psi^{(n)} \rangle \right ) - \lambda \langle 
\psi^{(n)}\rangle = 0\,, 
\end{equation} 
which is clearly the spectral truncation of the eigenvalue equation 
(\ref{gse}) for the continuous NLS system (\ref{nls1}). It follows,
therefore, that the mean--field  
predicted by our theory corresponds to a solitary wave solution of  
the NLS equation. On the other hand, the fluctuations $ \delta u_j$ and
$ \delta v_j$ are independant Gaussian variables with identical 
variance $ \frac{1}{\beta k_j^2}$.\\

The energy constraint (\ref{mfc}) imposes:

\begin{equation} \label{hameqn1} 
H^0  = \frac{n}{\beta} + H_n ( \langle u^{(n)} \rangle, \langle v^{(n)} 
\rangle)\,. 
\end{equation} 

where the term $ \frac{n}{\beta} $ reflects the equipartition of energy
among the $2n$ independant fluctuating modes.
The calculation of the total entropy is straightforward:

\begin{equation} \label{entmax} 
S(\rho^{(n)}) = C(n) + n \log \left ( \frac{L^2 [H^0 - H_n ( \langle 
u^{(n)} \rangle,\langle v^{(n)} \rangle)]}{n} \right ) \,. 
\end{equation} 
where $C(n) = n - \sum_{j=1}^n \log (j^2 \pi/2)$ depends 
only on the number of Fourier modes $n$. Thus, the key results follow the
maximization of the entropy: firstly the mean-field pair $(\langle u^{(n)} \rangle, 
\langle v^{(n)} \rangle)$ has in fact to realize the minimum possible value of 
$H_n$ over all fields $(u^{(n)},v^{(n)})$  that satisfy the  constraint  
$N_n(u^{(n)}, v^{(n)}) = N^0$. Moreover the excess energy not present in
the mean field is equally distributed among the fluctuating modes, with the 
inverse temperature:
\begin{equation} \label{betan2} 
\beta = \frac{n}{H^0 - H_n^*}\,. 
\end{equation} 
where $H_n^*$ is precisely this mimimum value of $H_n$ allowed by the particle 
number constraint $N_n = N^0$. We obtain also that the inverse temperature
scales linearly with the number of modes in the continuum limit.\\
The vanishing of fluctuations hypothesis is also verified if one computes
the contribution of the fluctuations to the number of particles. It reads:
\begin{equation} \label{partfluct} 
\frac{1}{2} \sum_{j=1}^n \left [ \langle (\delta u_j)^2 \rangle + 
\langle (\delta v_j)^2 \rangle  \right ]  
 =  \frac{H^0 - H_n^*}{n} \sum_{j=1}^n \frac{1}{k_j^2}   
 =  O(\frac{1}{n})\,,\;\; \mbox{as } n \rightarrow \infty\,. 
\end{equation} 

The limitation of our statistical description follows directly from Eq. 
(\ref{partfluct}). Indeed, for a fixed number of modes $n$, the coherent 
structure can only emerge if $ N \gg \frac{H^0 - H_n^*}{n}$. If this 
criterion is not satisfied, the condensation of the solution into the
coherent structure can even be broken. In fact, it is
still under debate whether such a coherent structure can emerge in the 
continuum limit when starting with a highly fluctuating state concentrated at
relatively small wavelengths\cite{picozzi}.
In this approach the fluctuation modes have been choosen as the $ e_j(x)$ 
Fourier functions. This assumption is in fact a direct consequence of the
factorization of the entropy, considering independent Fourier 
coefficients. Such an hypothesis is correct for defocusing homogenous NLS 
equation but is in general only valid for large wave number $k_j$.
Indeed, it is well known by linearizing eq. (\ref{nls1}) around a given 
state that the perturbation modes have to be found through the so called 
Bogoliubov theory. Although the full determination of the Bogoliubov modes is
needed to improve the statistics, we note that plane waves are a
good approximation of the fluctuation modes for large enough $k_j$ so that
our theory is always relevant to describe the statistics of the small 
wavelength perturbations.\\

Finally, regardless of these restrictions, our statistical equilibrium 
approach gives the following prediction for the particle number spectral 
density:
\begin{equation} \label{partspectra}  
\langle |\psi_j|^2 \rangle =  | \langle \psi_j \rangle |^2
 + \frac{H^0 - H_n^*}{n k_j^2}\,,
\end{equation} 
where we have used the identity $\psi_j = u_j + i v_j$.\\

\subsection{Numerical results}
\begin{figure}
\centerline{\epsfxsize=10truecm \epsfbox{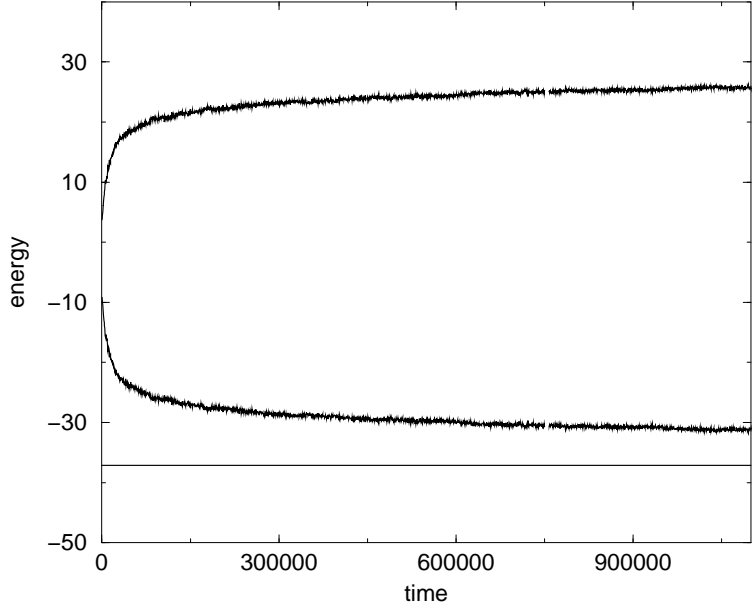}}   
\caption{\protect\small Kinetic (top curve) and potential (bottom curve) 
energies as function of time for the weakly focusing NLS. The straight line
below the potential energy corresponds to the potential energy of the
minimizer of the Hamiltonian for the particle number. The calculation
are made with $n=512$ discretization points and ensemble averaged is 
performed over $16$ different initial configurations.
\label{enerNLS}}
\end{figure}

In \cite{JJ00}, we discuss how numerical simulations compare with these
statistical predictions. Starting from any set of initial condition, one 
expects, following ergodic theory, that large time and large ensemble (over 
different initial conditions) averages will reproduce the statistical 
description of the system. Alternatively, one can choose to measure 
``quasi-static'' averaged quantities by measuring only few time unit 
averages and (if possible) large ensemble average. Such quasi--static 
description would converge to the sought after statistical regime for large enough
integration time of the dynamics where the transitory effects due to the 
initial conditions can be
ignored. We address here the weakly focusing nonlinearity $f(|\psi|^2)
=|\psi|$ already used for figure (\ref{evolu}). The transient dynamics can
be estimated by plotting the two contributions to the conserved total energy.
Figure (\ref{enerNLS}) shows over a large period the
evolution of the kinetic and the potential energy, averaged over $16$ 
slightly noisy homogenous states. Fast but small oscillations of the energies
around a smooth evolution account for the rapid fluctuation modes.
The kinetic (potential) energy is then found to
increase (decrease) as time goes on, indicating the transfer of energy 
towards smaller scale.
Moreover, after a short transient (until 
$t=50000$ time units) corresponding to the formation of peaks and to
their coalescence into one single structure, a slow quasistatic regime
is observed as described in figure (\ref{evolu}). This slow evolution 
witnesses the convergence of the coherent structure to the Hamiltonian 
minimizer. The straight line below the potential energy 
indicates the potential energy of the minimizer for the total 
particle number $N_0$. The potential energy approaches a constant value close
to this bound, the difference being due to the finite particle number which
still remains in the fluctuations. This quasistatic dynamics seems in fact to 
saturate for this $n$-modes calculations at large time ($t\ge 900000$).
Indeed, such convergence is observed on the mean particle number spectra 
averaged over few time units and $16$ initial conditions, which shows 
stationary properties thereafter. 

\begin{figure}
\centerline{  \epsfxsize=10truecm \epsfbox{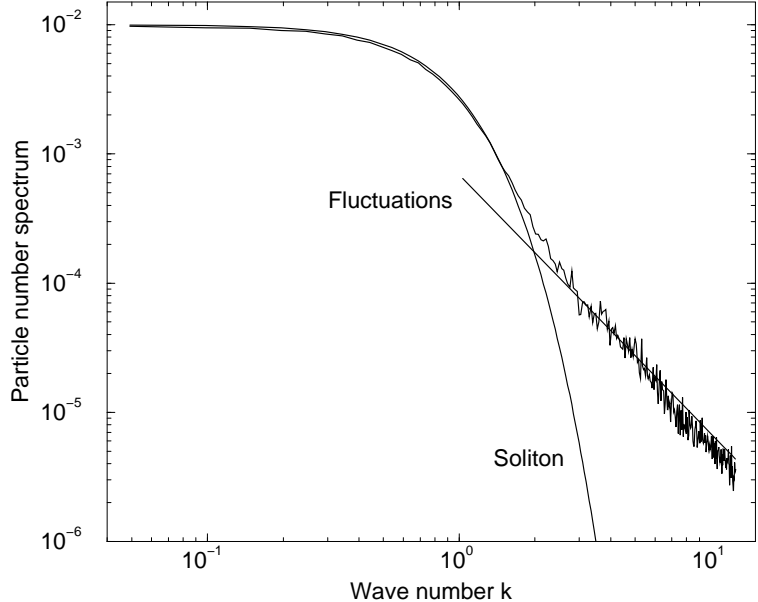}}   
\caption{\protect\small Particle number spectral density
 $ |\psi_k|^2$ as a 
function of $k$ for $t=1.1\times10^6$ 
unit time (upper curve). The lower curve  
(smooth one) is the particle number spectral
density  for the solitary wave  that
contains all the particles of the 
system. The straight line drawn for large $k$ corresponds to
the statistical prediction (\ref{partspectra}) for
the spectral density for large wavenumbers. The numerical simulation has 
been performed with $n=512$, $dx=0.25$, $N^0=20.48$ and $H^0=-5.46$. 
\label{specnor}}
\end{figure}

Figure (\ref{specnor}) shows this particle number spectral density. 
The spectral density of the solitary solution which contains the total initial 
particle number is drawn (smooth line) for comparison.
The equipartition for large wavenumber deduced from the theory is
also marked, corresponding to a straight line in the log-log plot. Very good
quantitative agreements are found between the theoretical predictions and
the numerics, for the coherent structure at large scale as well as for the 
$1/k^2$ spectrum for small scale.

\begin{figure}
\centerline{  \epsfxsize=10truecm \epsfbox{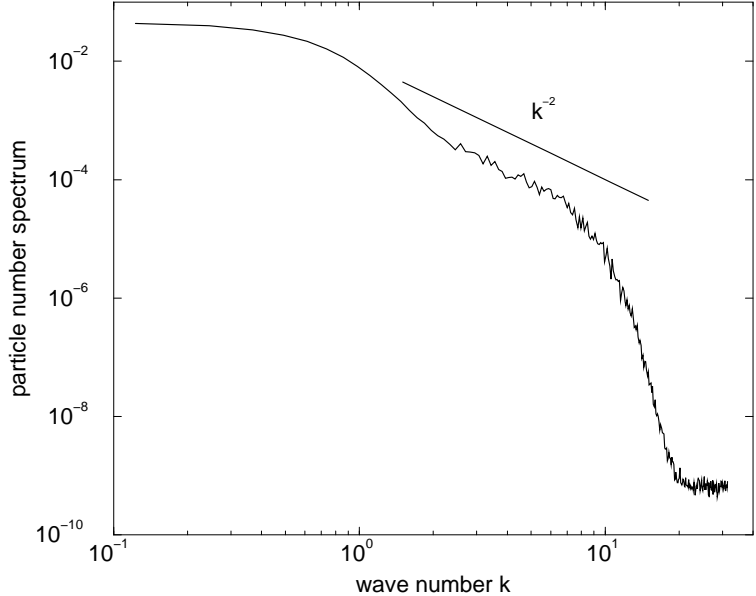}}   
\caption{\protect\small The particle number spectral density for $n=512$ and 
$dx=0.1$ (thus $L=51.2$) at unit time $t=5 \cdot 10^5$. The coherent soliton
structure already accounts for almost the entire  number of particles
of the system,  but the system has not yet reached  statistical equilibrium. 
The initial noise level is still present for large wavenumbers ($ k \ge 20$), 
while at smaller wavenumbers, one can recognize both the soliton-like 
structure and a fluctuation spectrum following approximately a $k^{-2}$
law. The spectrum has been obtained by an ensemble average over 16 initial 
conditions and a time average over the final 10 unit times.
\label{inter}} 
\end{figure}

Before this stationary regime is attained, one can ask how the system 
finds its road to the statistical equilibrium starting from a given initial
conditions (taken here as homogenous states perturbed by a slight random 
noise). Since the dynamics after the coalescence regime is quasistatic, 
converging slowly towards the statistical equilibrium, one can follow this 
evolution using the average of the particle number spectral density at time
$t$. Here the mean value is taken over the ensemble of $16$ initial 
conditions and also by time averaging over few time units only. Thus the 
average smoothes the fast variations while the evolutions due to the 
quasistatic dynamics can be neglected.
Figure (\ref{inter}) shows the particle spectrum for such an 
intermediate time $t_i$. The statistical equilibrium has not been reached yet, 
but the solution has already converged into a single solitary wave surrounded
by fluctuations. The spectrum at long wavelengths accounts again for the 
soliton-like coherent structure containing already almost all the total
particles. On the other hand at small scales, the fluctuation modes 
already exhibit the $1/k^2$ law but only for wavenumbers smaller than a well
defined wavenumber $ K(t_i)$. For higher wavenumbers, the amplitudes 
of the fluctuation modes are at the level of the initial noise. Somehow, 
the system acts as if at each time $t$ the statistical equilibrium 
(\ref{partspectra}) were satisfied only for scales larger than $\lambda(t)=
2 \pi/ K(t)$. As time goes on, the front $ k=K(t)$ advances in the momentum
space as the  ``quasistatic'' equilibrium invades smaller and smaller 
lengthscales.
For a finite number of modes, this slow dynamics saturates when all 
avalaible modes are reached by the equilibrium. Thermal equilibrium between 
the fluctuations has ben attained.
The front dynamics is found
to obey the scaling $ K(t) \simeq t^{1/4}$ using high derivative moments of
the wavefunction $\psi$. It is then tempting to generalize these results to 
the continuum limit $ n \rightarrow \infty $. In that case, we expect that
at any time the dynamics is made of a solitary coherent structure containing
most of the particle number in quasi-statistical equilibrium with a finite
number of fluctuating modes. As time goes on, the number of modes at
equilibrium increases and the effective temperature due to the equipartition 
of the energy among the modes decreases. Thus the contribution of the 
fluctuations to the total mass decreases providing an inverse transfer of
mass from the fluctuations to the soliton-like structure. The $t \rightarrow 
\infty$ limit corresponds to {\it the} coherent structure
containing {\it all} the mass of the system while the excess energy is
distributed over an infinite number of modes of zero amplitude!

One may have argued that the prediction of the equipartition of the energy 
among the modes surrounding a {\it condensate} structures was obvious since 
some sort of thermal equilibrium was assumed. In fact, it is noteworthy to 
observe that the dynamics of (\ref{nls1}) reaches in fact a self-thermalized 
state through the dependance of the temperature with the number $n$ of 
available modes.  The singular limit $n \rightarrow \infty$ can then be 
interpreted as a semi-classical example of the Rayleigh-Jeans 
paradox\cite{Pomeau}.

\section{Wave turbulence in BEC}\label{secBEC}

We outline here the conclusions of the previous section for a model of BEC. 
For sake of simplicity and to allow long numerical simulations we 
will again restrict our study to one spatial dimension. In dimensionless
units, the dynamics of a BEC trapped in a harmonic potential 
reads\cite{DGPS99}:

\begin{equation} \label{bec} 
i \partial_t \psi= \left(-\frac{1}{2}\Delta  + V_{ext}(x)+ 
|\psi|^2 \right) \psi
\end{equation} 

where the additional harmonic potential $ V_{ext}(x)=\frac{1}{2}\Omega^2 x^2$ 
describes the external potential used to confine the particles in
an atomic trap. The usual defocusing nonlinearity corresponding to 
repulsive particle interactionsis used. The 
integrability of the 1D NLS equation is broken here due to the presence of
the external potential $V_{ext}$. The structure of the equation is clearly
equivalent to the general NLS system. The number of particles:

$$ N=\int |\psi|^2dx  $$

is conserved as well as the Hamiltonian

\begin{equation}
H = \frac{1}{2} \int\left ( |{\bf \nabla} \psi|^2+\Omega^2 x^2 |\psi|^2  +|\psi|^4 \right)dx \,.
\label{hambec}
\end{equation}

To the kinetic energy $ |{\bf \nabla} \psi|^2 $ and the nonlinear potential
term $|\psi|^4$ is added the contribution of the external potential $ 
\Omega^2 x^2 |\psi|^2 $.
The solitary solutions $ \phi(x) e^{-\imath \mu t} $ which minimize the Hamiltonian
for a given number of particles $N$ are solutions of:

\begin{equation} \label{gbec} 
\mu \phi=-\frac{1}{2}\phi_{xx}   + \frac{1}{2} \Omega^2 x^2 \phi + \phi^3= 0\,. 
\end{equation} 

\begin{figure}
\centerline{  \epsfxsize=10truecm \epsfbox{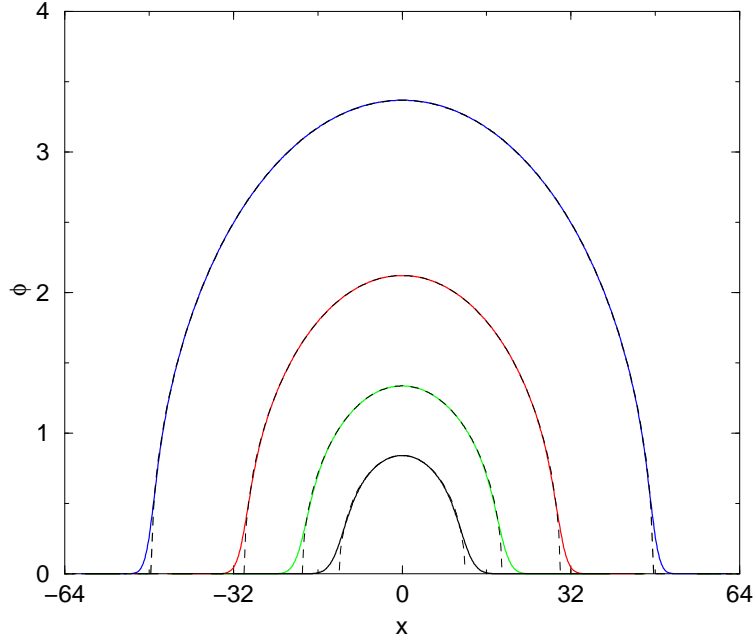}}   
\caption{\protect\small Solutions $\phi(x)$ of Eq. (\ref{gbec}) for 
$\Omega=0.1$ and different particle numbers $N=11.5$, $45$, $180$ and $720$
(as expected, the smaller $N$ the smaller $\phi$ on the figure). For each
solution the dashed line show the Thomas-Fermi approximation $ \phi_{TF}$. 
\label{solitons}}
\end{figure}

Such solutions can be obtained numerically and are well described for large
enough $N$ by the so-called Thomas-Fermi approximation. Neglecting the
kinetic term the Thomas-Fermi solution $\phi_{TF}$ only balances
the nonlinearity and the external potential:

$$ \phi_{TF}(x)= \sqrt{\mu-\frac{1}{2} \Omega^2 x^2} $$

for $ |x| \le \frac{\sqrt{2 \mu}}{\Omega}$ and $ \phi_{TF}(x)=0$ elsewhere.
By a straightforward integration the chemical potential $\mu_{TF}$ satisfies:

$$ \mu_{TF}= \frac{1}{2} \left(\frac{3N \Omega}{2} \right)^{\frac{2}{3}} $$

where $N$ is the total particle number.
The radius $R_{TF}$ of the condensate in the Thomas-Fermi 
approximation is found: $R_{TF}=\left(3N/(2\Omega^2) \right)^{1/3}$. Figure 
(\ref{solitons}) shows solitary solutions of Eq. (\ref{gbec}) for different 
values of $N$ together with the corresponding Thomas-Fermi solution. 
We observe that the 
Thomas-Fermi approximation is very good except in a small boundary layer near
$x=R_{TF}$. The potential $ \mu$ is also found numerically very close to 
$\mu_{TF}$. The solutions were calculated on the periodic box $ x\in [-64,64]$
of length $L=128$ unit length, with $\Omega=0.1$, that we will keep constant 
throughall later on.

Perturbations around this ground state are studied to determine the 
eigenfrequencies of the fluctuation modes. Following Bogoliubov approach, we 
are seeking solutions for the weak perturbation regime:

$$ \psi(x,t)= \left( \phi(x)+u(x)e^{-i\omega t}+v(x)e^{i\omega t} \right) e^{-i\mu t} $$

At first order in the complex functions $u$ and $v$, we obtain from (\ref{bec})
the coupled equations:

\begin{eqnarray}
\omega u(x)=\left(-\frac12 \partial_{xx}+\frac12 \Omega^2x^2-\mu+2 \phi(x)^2
\right) u(x)+\phi(x)^2 v(x) \label{bogu} \\
-\omega v(x)=\left(-\frac12 \partial_{xx}+\frac12 \Omega^2x^2-\mu+2 \phi(x)^2
\right) v(x)+\phi(x)^2 u(x) \label{bogv}
\end{eqnarray}

This linear system can be solved numerically to determine the 
excitation spectrum\cite{dodd}. However, as for the NLS system described above,
high frequency modes are well approximated by plane waves $ u(x),\,v(x)
\propto e^{ikx}$ for small wavelengths such that $ k^2 \gg
\mu \simeq \mu_{TF}$. We retrieve then the well-known Bogoliubov relation for 
high wavenumber dispersive waves 
$$ \omega= \frac12 k^2 \,. $$

{\it A priori}, the statistical approach developped above applies to this
specific NLS system (\ref{bec}). Thus, starting from any initial condition, 
we should observe the formation of a coherent structure solution of 
(\ref{gbec}) containing almost all the particle number of the system, in the
midst of wave fluctuations. For large enough time, one should observe also the 
statistical equipartition of the excess energy between all the 
avalaible modes for a finite-$n$ grid point calculation. In 
particular, we should observe again a 
$ 1/k^2$ particle number spectrum for large wavenumbers.
To illustrate this dynamics, we initiate the dynamics with the ground state of
a condensate of noninteracting bosons for $N=180$.
When the nonlinear term is neglected, the ground state of the 
linear Schr\"odinger equation is described by a gaussian distribution of
the particle around the trap center:

$$ \psi_i(x,t)=\sqrt{N} \left(\frac{\Omega}{\pi}\right)^\frac{1}{4} e^{-\frac{\Omega x^2}{2}}e^{-\imath \mu t} $$

\begin{figure}
\centerline{  a) \epsfxsize=7truecm \epsfbox{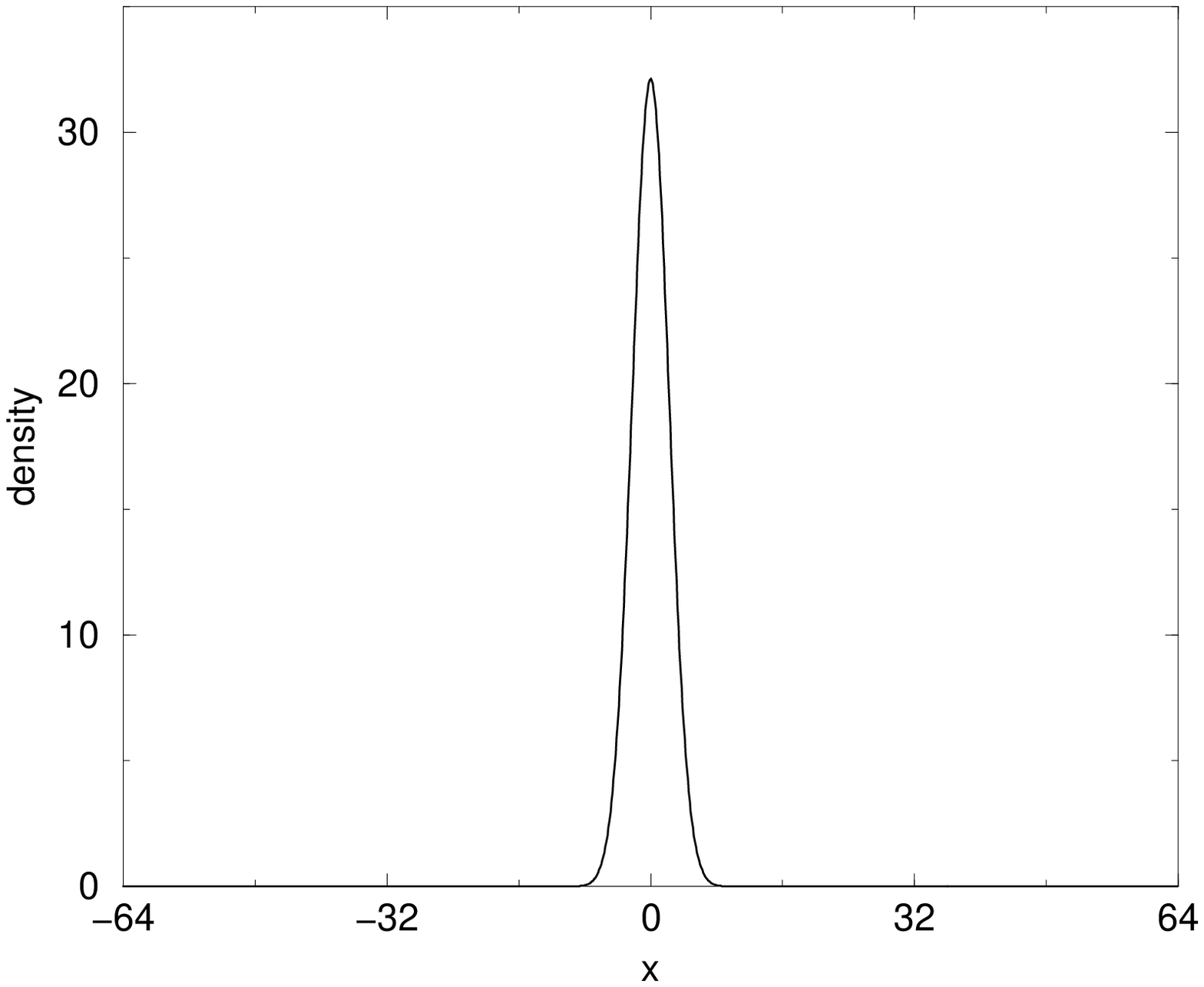} b) \epsfxsize=7truecm \epsfbox{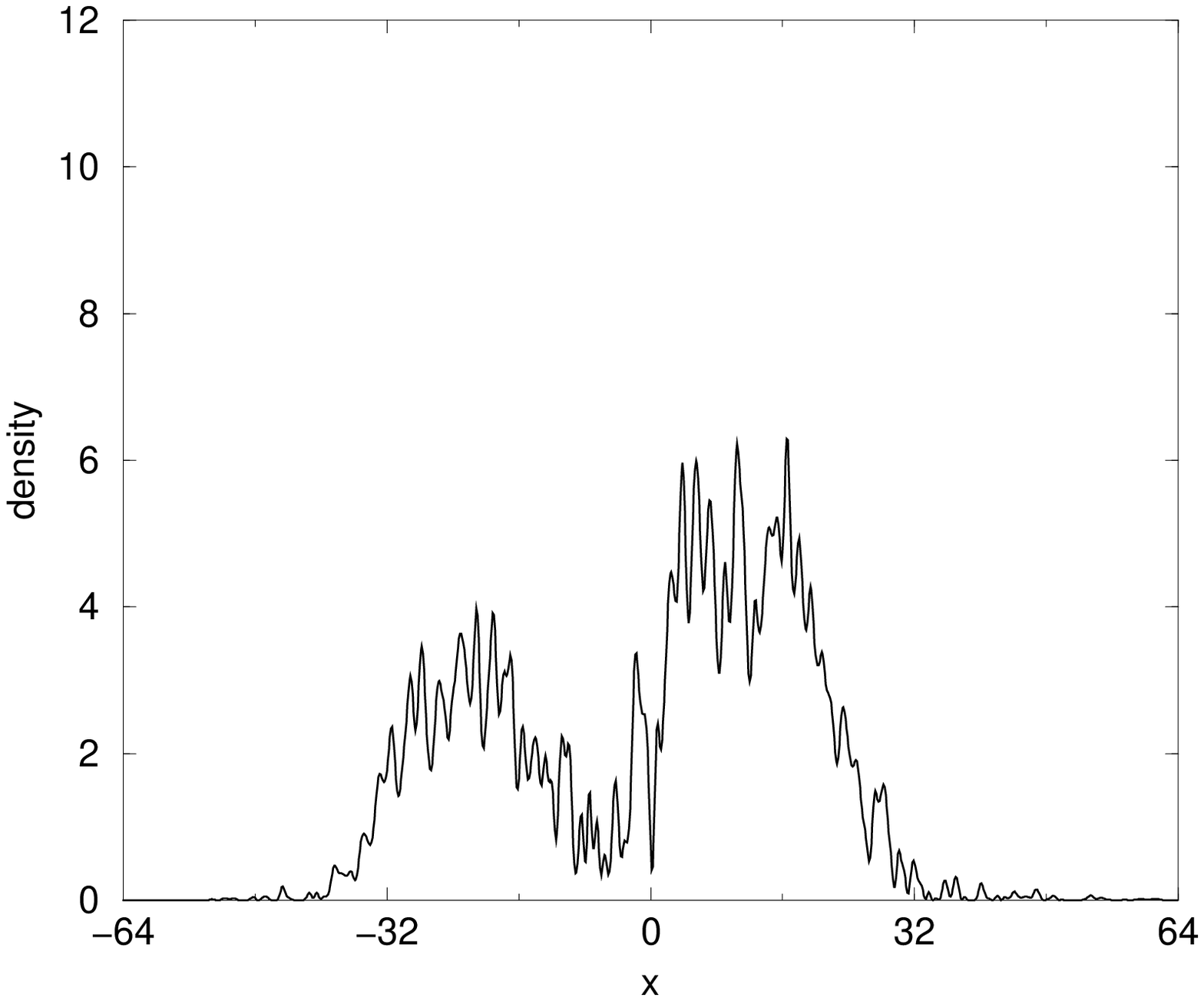}}
\centerline{ c)\epsfxsize=7truecm \epsfbox{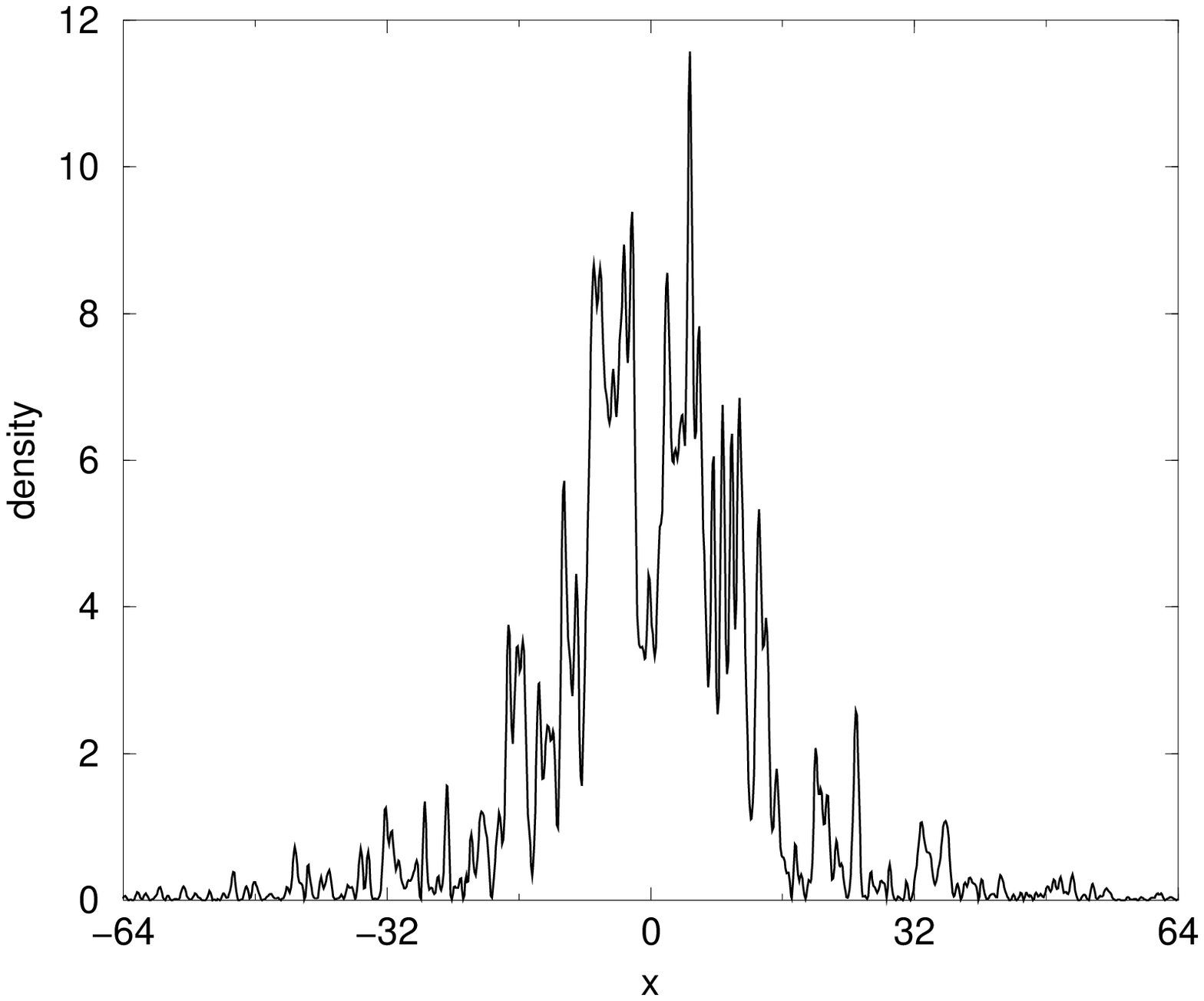} d) \epsfxsize=7truecm \epsfbox{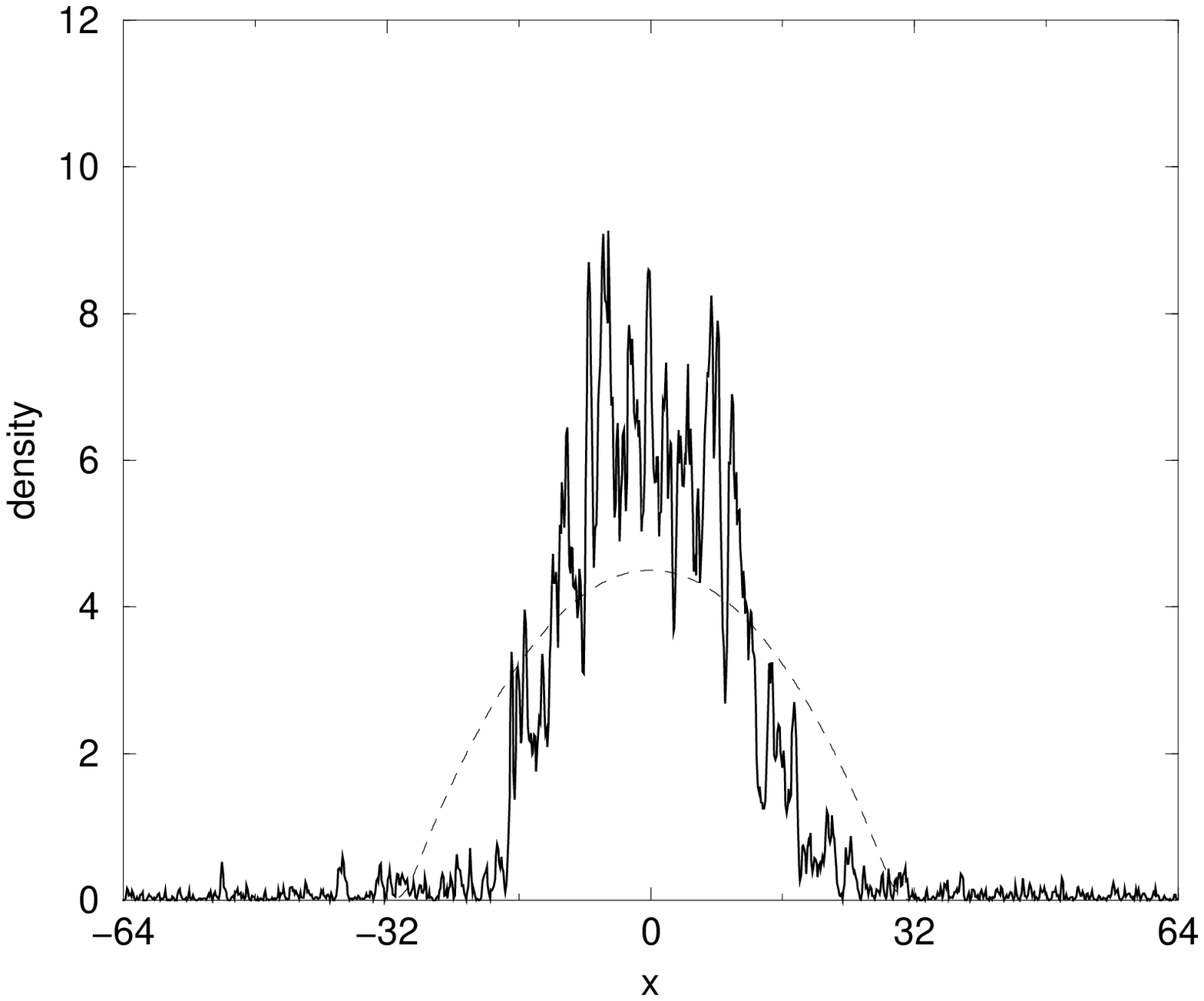}} 
\caption{\protect\small Particle number density $ |\psi(x,t)|^2$ for different
times: a) initial gaussian profile at $t=0$, b) $t=4000$, c) $t=34000$ and
d) $t=250000$ units time. The coherent structures containing all the 
particles is shown as dashed lines on figure d).
\label{snapbec}}
\end{figure}

The gaussian width $\sigma=1/\sqrt{\Omega}$ defining the typical radius of the 
noninteracting gas.  Numerical simulations are typically
made on a regular grid of $512$ or $1024$ points, using a Yoshida 
pseudospectral 
splitting scheme\cite{schatz} which is $4$th order in time. Mass and energy 
are conserved to within $10^{-10}$ and $10^{-5}$ relative errors respectively.
At $t=0$ we start the dynamics of Eq. (\ref{bec}) with:

$$ \psi(x,0)=\psi_i(x,0)+\eta(x) $$

where $ \eta(x)$ is a white noise of very small amplitude ($10^{-5}$), taken 
to break the $ x \rightarrow -x$ symetry of the system. Figure (\ref{snapbec})
show snapshots of the solution for different times.

We observe roughly the expected dynamics where the coherent structure 
emerges from fluctuations. However, one can notice that the fluctuation
amplitudes are much higher here than for the simulations shown in the
previous section, so that the comparison between the solution at large times 
and the coherent structure (figure \ref{snapbec} d)) is only partially 
satisfied. This high level of fluctuations is due to the large
difference of energy between the initial condition ($H=564.25$) and the 
coherent structure $\phi(x)$ minimizing the Hamiltonian ($H^*=486.15$).
Moreover, since the initial gaussian width ($\sigma=3.16$) is smaller than the 
Thomas-Fermi radius ($R_{TF}=30$) of the coherent structure, we observe
in the numerics numerous large radius oscillations around $R_{TF}$ before 
the solution stabilizes. Such radius oscilations can be noticed between
figures (\ref{snapbec} a), b), c)).

\begin{figure}
\centerline{\epsfxsize=10truecm \epsfbox{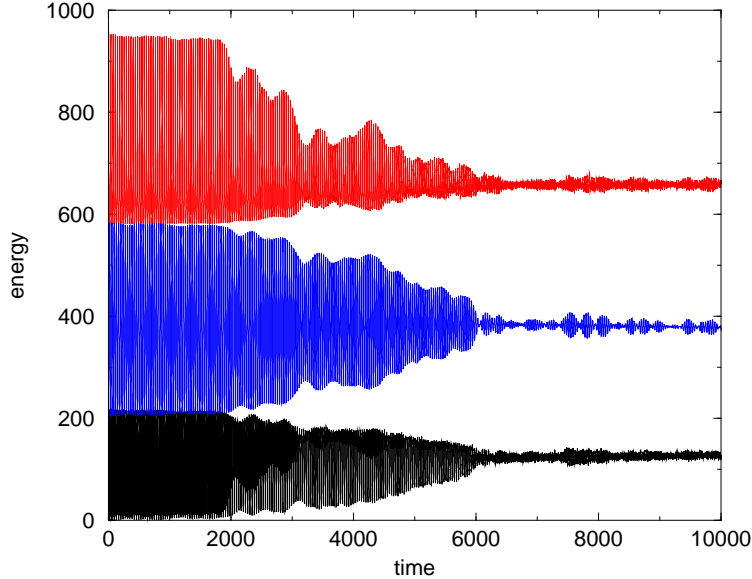}}   
\caption{\protect\small Different contributions to the total constant
energy as functions of time for short times. Kinetic energy (black curve),
external potential contribution (red curve) and nonlinear one (blue curve)
have been also vertically translated for vizualisation. The calculation
is made with $n=512$ discretization points.
\label{ener512}}
\end{figure}

\begin{figure}
\centerline{\epsfxsize=10truecm \epsfbox{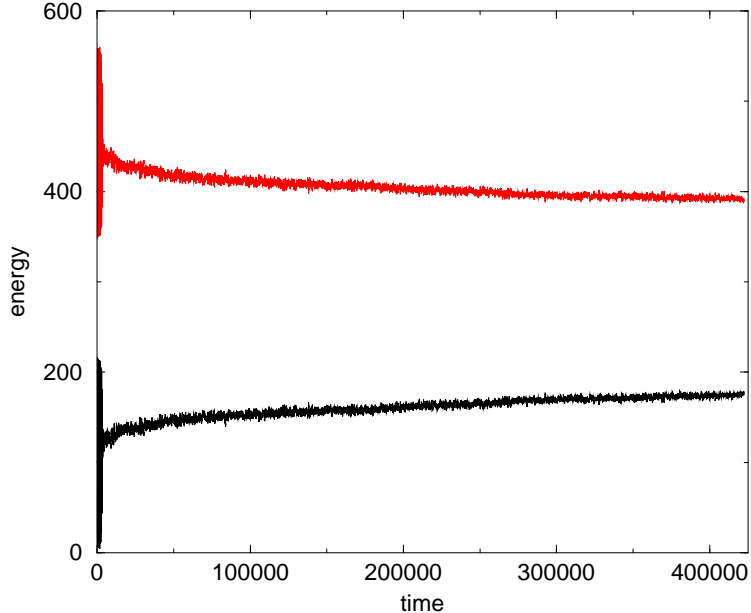}}   
\caption{\protect\small Kinetic energy (bottom black curve),
sum of the external potential and of the nonlinear contribution (top red curve)
as function of time. The calculations
are made here with $n=1024$ discretization points.
\label{ener1024}}
\end{figure}

These oscillations and their 
``effective'' damping are even better seen on figure (\ref{ener512}) where
the different (kinetic, external potential and nonlinear) contributions to 
the total energy are shown at short time. Recall that the sum of these 
three contributions is constant throughout the dynamics.
The amplitude of these oscillations are rapidly decreasing after few thousand
time units. Thereafter, only small oscillations are observed around slowly
varying energy contributions. This slow dynamics is shown 
on figure (\ref{ener1024}) where again kinetic (bottom curve), and the sum of 
the external and the potential energies (top curve) are drawn on a much 
larger time scale.
After the short transient, where fluctuations are important, we observe a 
quasistatic dynamics consisting of small rapid fluctuations around a slow
variation of the energy contribution. As expected, the kinetic contribution
increases, while obviously the other contributions decrease. 
This slow dynamics indicates as for the previous focusing NLS 
system the transfer of energy towards smaller and smaller scales as time 
goes on.

\begin{figure}
\centerline{\epsfxsize=10truecm \epsfbox{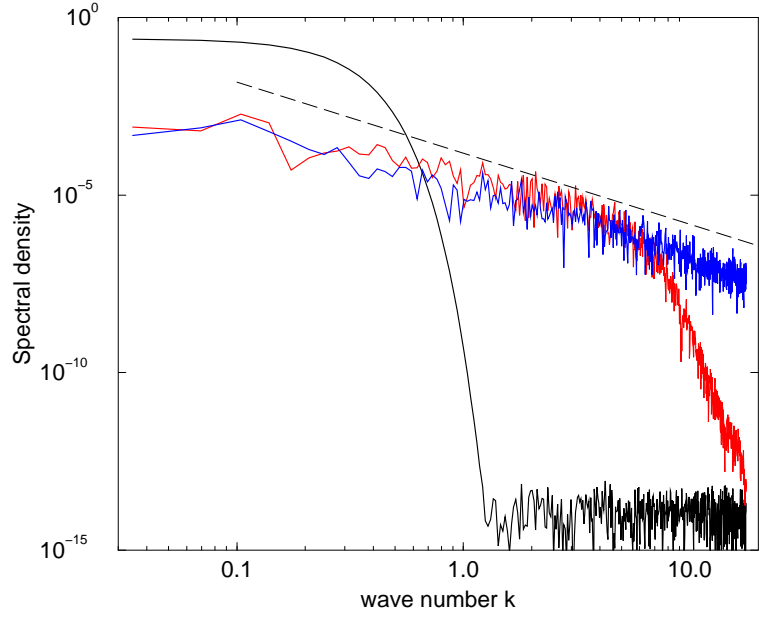}}   
\caption{\protect\small Particle number spectral density of the fluctuations
around the solution $\phi(x)^e{-i\mu t}$ at different stage of the 
dynamics on a log-log scale. From bottom to top at high wave number, we have the spectrum
of the initial condition  $t=0$ (black curve), the spectrum at $t=16900$ 
time units (red curve) and at $t=422000$ (blue curve). The dashed line 
represents the spectral density for large wavenumber as predicted by our 
theory. The number of grid points for calculation is $n=1024$ here.
\label{spec}}
\end{figure}

The cascade-like transfer can also be seen on figure (\ref{spec}) where the 
particle number spectral density of the fluctuations are shown at
different instants of the dynamics. This spectral density is obtained by 
substracting the coherent structure $ \phi e^{-i\mu t}$ from the solution
$\psi(x,t)$. The spectra are obtained by time averaging over one unit of
time, except for the initial condition spectrum. One can see the
evolution of the spectrum from the initial gaussian distribution to the 
predicted spectrum at $t=422000$. The dashed line represents the quantitative
prediction $\langle |\psi_j|^2 \rangle =  \frac{H^0 - H_n^*}{n k_j^2}$, based
on $ H_n^*$ calculated with the soliton-like solution containing all the
particles ($N=180$). Such a stationary spectrum corresponding to the statistical 
equilibrium distribution was actually mostly attained since $t=200000$ 
time units. The curve at $t=16900$ indicates that the $1/k^2$ spectrum is
fist reached at long wavelengths and invades smaller and smaller lengthscales
as time goes on, as already observed for the focusing NLS of the
previous section. 

\begin{figure}
\centerline{\epsfxsize=10truecm \epsfbox{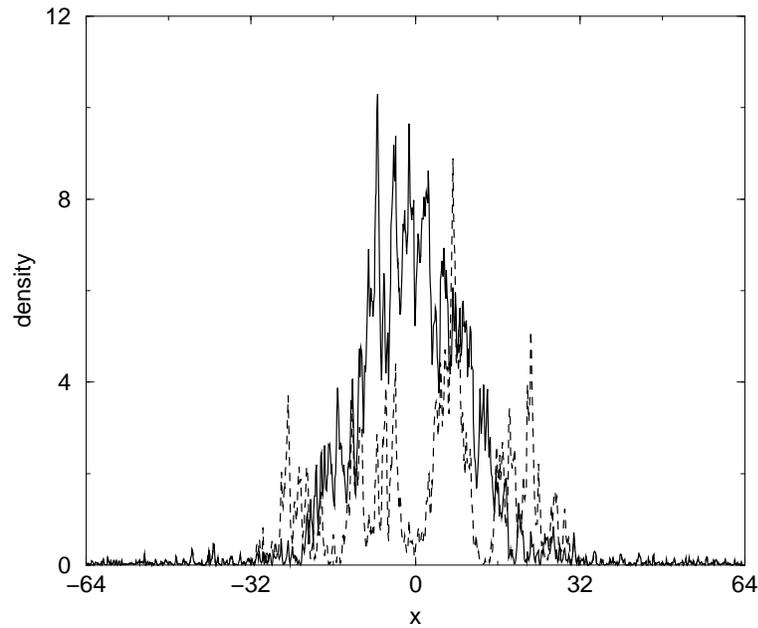}}   
\caption{\protect\small Particle density $|\psi(x,t)|^2$ (solid line) of the 
solution of the dynamics (\ref{bec}) and the density of the fluctuation field
(dashed curve) at $t=422000$.
\label{fluctuations}}
\end{figure}

A difference between the predicted spectrum amplitude at high wave numbers 
and the numerical results is observed, while the agreement for the focusing 
NLS in the
previous section was much better. We understand such discrepancy by a higher
level of fluctuations in the BEC system. The energy difference between the
initial condition and the ground state coherent structure containing all the
particles is $\Delta H =78$ here while it was only a few units for the focusing 
NLS case. Thus the amplitude of the fluctuations are the same order as
the coherent structure, as observed on figure (\ref{fluctuations}) where,
both the density of the numerical simulations and the 
density of the fluctuation field deduced after sustraction of the expected 
coherent structures are shown at $t=422000$. Here we are close to the
limitation of our approach since the particle number of the fluctuations 
field is close to the total particle number as discussed previously.

This finite mode self-thermalization of the NLS-like equations has been used
recently to model finite temperature BEC\cite{davis02,davis03,castin03}. Distributing fluctuation energy equally
among the Bogoliubov modes, the authors ``mimic'' numerically a thermal
system at equilibrium with only the Gross-Pitaevski\~{\i} equation. 
Although the cutoff at high wavenumbers of the 
black-body-like catastrophe is needed, we want to emphasize here that this 
approach is inconsistent since no convergence of this regularization in the 
continuum limit is proposed. A consistent regularization would have to 
integrate a quantum description of the dynamics.
Our work here, without solving this crucial 
question, can be seen as an attempt to describe the dynamics of a classical field.\\

\section{Conclusion}

We have presented here the asymptotic dynamics of different nonlinear 
differential equations used to model superfluids and BEC. The description of
the statistical properties of the solution for large times when a 
quasi-static regime is reached has been developped for a finite number of 
modes truncation. Seeking a stationary probability density function 
we were led to
consider the maximum entropy density where all the mass is 
contained in the coherent structure minimizing the Hamiltonian in the 
continuum limit. Thus, the large scale solutions of the dynamics for a finite
number of modes are found by this approach to behave like this coherent 
structure immerged in a sea of energy equipartitioned fluctuations.
Long time numerical simulations show then very good agreement 
with the predictions both for a weakly focusing NLS equation and for a 1D
version of the BEC model. The road towards these statistically stationary 
states gives a consistent scenario of the continuum limit of the dynamics.
It corresponds to a quasistatic dynamics where the number of 
fluctuation modes at equilibrium increase with time. The kinetics of
the advance towards smaller and smaller scales has been studied for a particular case of NLS dynamics and remains
to be fully explored in general. 

\section*{Acknowledgements}
It is my pleasure to thank Richard Jordan with whom this work was initiated.
I want also to acknowledge the editors of this special issue for encouraging 
me to come back to this exciting subject.

\end{document}